\documentclass[aps,nofootinbib,preprintnumbers,twocolumn,superscriptaddress,prd]{revtex4-1}

\usepackage{amsmath}
\usepackage{amssymb}
\usepackage{slashed}
\usepackage{epsfig}
\usepackage{bbm,bm}
\usepackage{hyperref}
\usepackage{color}
\usepackage{comment}
\usepackage{amsfonts}
\usepackage{dsfont}

\usepackage{multirow}

\def\bea{\begin{eqnarray}} 
\def\eea{\end{eqnarray}}
\def\be{\begin{equation}} 
\def\ee{\end{equation}} 
\def\ba{\begin{array}}
\def\ea{\end{array}} 
\def\nn{\nonumber}

\def\be{\begin{equation}}
\def\ee{\end{equation}}
\def\bea{\begin{eqnarray}}
\def\eea{\end{eqnarray}}

\def\nn{\nonumber}

\usepackage{xcolor,colortbl}

\definecolor{Gray}{gray}{0.9}
\newcolumntype{a}{>{\columncolor{Gray}}c}

\let\oldtitle\title
\renewcommand{\title}[1]{\oldtitle{\color{blue}{#1}}}

\begin{document}

\title{
Symmetry and universality of multi-field interactions in $6\!-\!\epsilon$ dimensions
}

%

\author{A.\ Codello}
\email{a.codello@gmail.com}
\affiliation{Instituto de F\'isica, Faculdad de Ingenier\'ia, Universidad de la Rep\'ublica, 11000 Montevideo, Uruguay}

\author{M.\ Safari}
\email{safari@bo.infn.it}
\affiliation{Romanian Institute of Science and Technology, 
Str.~Virgil Fulicea  3, 400022 Cluj-Napoca, Rom\^ania}

\author{G.\ P.\ Vacca}
\email{vacca@bo.infn.it}
\affiliation{INFN - Sezione di Bologna, via Irnerio 46, I-40126 Bologna, Italy}

\author{O.\ Zanusso}
\email{omar.zanusso@unipi.it}
\affiliation{Universit\`a di Pisa and INFN - Sezione di Pisa, Largo Bruno Pontecorvo 3, I-56127 Pisa, Italy}

\begin{abstract}
We outline a general strategy developed for the analysis of critical models,
which we apply to obtain a heuristic classification of all universality classes
with up to three field-theoretical scalar order parameters in $d=6\!-\!\epsilon$ dimensions.
As expected by the paradigm of universality, each class is uniquely characterized by its symmetry group
and by a set of its scaling properties, neither of which are built-in by the formalism
but instead emerge nontrivially as outputs of our computations.
For three fields, we find several solutions mostly with discrete symmetries. These are nontrivial conformal field theory candidates in less than six dimensions, one of which is a new perturbatively unitary critical model.
\end{abstract}

\pacs{}

\maketitle

\section{Introduction}

Quantum and statistical field theories are the most powerful theoretical
tools to investigate the physics of critical and almost-critical systems.
The predictive power of field theory is astonishing because it works equally well
in diametrically opposed domains of physics:
high energy physics of particles with fundamental interactions and low energy physics
of systems subject to statistical fluctuations.
An explanation for such broad applicability range can be argued on the basis
of three powerful pillars of modern physics: scaling, universality, and symmetry.
However, it remains fundamentally mysterious how nature's degrees of freedom (fields)
arrange themselves into interactions.
One can expect that critical and almost critical theories are allowed to exhibit only specific symmetries in their field dynamics.
A fitting analogy would be to see the unconstrained space of all possible theories as a tumultuous sea which we know very little of, 
but in which the 
critical points act as lighthouses helping us in the process of charting.
Needless to say, a chart of all possible field theories would be extremely desirable and have strong physical implications as it could guide experimental and numerical investigations of critical systems to find physical realizations of all the
possible critical points.
On the basis of traditional arguments of scaling and universality,
we expect that any found critical theory could be realized in nature,
either as some infrared effective models at large distances, but even as fundamental models, i.e.\ ultraviolet complete field theories.
This paper discusses a powerful and completely general method to unveil this pattern.
To illustrate the method we give as first application the heuristic classification
of all possible critical theories with cubic interactions and up to three order parameters in $d=6-\epsilon$ dimensions. 
We have chosen to embark on this example because it reveals some unexpected and previously unknown critical theories.

\section{Preliminaries}
We stress that the most ambitious part of this approach is that we want to let symmetry emerge
from the condition of criticality, rather than input a certain symmetry content \emph{a priori}. 
Given a specific field content it is rather straightforward to write down all possible interactions,
but in the absence of a constraint dictated by symmetry the number of possible interactions grows fast with the number of field theoretical degrees of freedom. 
This problem can be studied in a fully consistent way using renormalization group (RG) methods~\cite{Wilson:1971bg, Wilson:1973jj}
and the $\epsilon$-expansion~\cite{Brezin:1973jt} as main paradigms to carry on our investigations. 
A critical theory is seen as a scale invariant fixed point of the RG flow
and critical parameters can be determined order-by-order in $\epsilon$. 
Scale invariance is often lifted to conformal symmetry; therefore, we must make sure that our tools integrate well with conformal field theory (CFT) methods~\cite{ElShowk:2012ht, Rychkov:2015naa, Codello:2017qek, Codello:2018nbe, Gopakumar:2016wkt,Alday:2016njk, Gliozzi:2016ysv}. 
In this work we adopt functional perturbative RG methods~\cite{Jack:1982sr,Codello:2017hhh,Martini:2018ska} because they suit well the analysis of multi-scalar QFTs with no symmetry imposed a priori.
The complexity of the unconstrained algebraic problem (because of the absence of symmetries) is reduced taking advantage of elementary group theory considerations.

In past RG investigations of theories with quartic interactions in $d=4-\epsilon$, the trace condition \cite{Brezin:1973jt,Pelissetto:2000ek} for critical potentials was assumed.
This leads to complete degeneracy of the field anomalous dimensions and constrains to one the number of control parameters, therefore reducing the possible number
of critical theories~\cite{Wallace:1975ez, Zia:1974nv, Brezin:1973jt, Michel:1983in, TMTB,Rychkov:2018vya}.
A systematic search for all possible theories with $N=2$ fields has been done in~\cite{Osborn:2017ucf}, but a general study with $N\geq 2$ has yet to come.
In the following, we concentrate on systems with cubic interactions and upper critical dimension $d_c=6$, which could be seen as generalizations of the Lee-Yang model~\cite{Fisher:1978pf}.
Multi-field theories of this type, exhibiting a singlet and a vector under $O(N)$ symmetry, have already been studied in~\cite{Fei:2014yja} showing that perturbatively unitary models might emerge above four dimensions (a theory is perturbatively unitary 
if all operator scaling dimensions are within the unitarity bound \cite{Fei:2014yja}).
We undertake the next step and perform a complete perturbative analysis of cubic theories with up to $N=3$ real scalar fields. 
The results reveal a large unexplored theory space which includes some critical theories that can correspond to new universality classes in lower dimensions, including $d=2,3$ and $4$.


\section{Beta functions}


We work with a functional perturbative formalism~\cite{Jack:1982sr,ODwyer:2007brp,Codello:2017hhh,Osborn:2017ucf}, 
which more conveniently encodes at once information about the critical theory and its deformations induced by relevant operators.
At one-loop level (see appendix~\ref{sect:appendix-betas} for the next to leading order), the functional beta for the dimensionless potential $v(\phi)$ in $d=6-\epsilon$ is 
\be 
\beta_v = -d v +\frac{d-2}{2}\phi_i v_i + \phi_i \gamma_{ij} v_j - \frac{2}{3} v_{ij}v _{jk} v_{ki} \,,
\ee
in which Latin indices such as $i$ on $v$ run over the number of flavors ($i=1,\dots,N$) and denote derivatives w.r.t.\ the corresponding field component $\phi_i$. In our conventions, we have also rescaled the potential as $v \to 2(4\pi)^{3/2} v$.
We will restrict ourselves here to scalar theories with at most three flavors ($N=3$). No symmetry is imposed on the model and we therefore consider the most general parametrization of the cubic interactions given in terms of the dimensionless potential
\be  \label{v}
v(\phi) = {\textstyle{\frac{1}{3!}}} \lambda_{ijk}\,\phi_i\phi_i\phi_k\,.
\ee
In terms of the classically marginal couplings $\lambda_{ijk}$, the leading order (LO) one-loop anomalous dimension matrix is given by 
\be 
\gamma_{ij} = \frac{1}{3} \lambda_{iab} \lambda_{jab} \,.
\ee
Its eigenvalues evaluated at the fixed point are denoted by $\gamma_i$.
Taking into account the contribution from the anomalous dimensions, the couplings $ \lambda_{ijk} $ flow according to the following beta functions:
\be \label{beta-3}
\beta_{ijk}
= -\frac{1}{2}\epsilon\, \lambda_{ijk} + \lambda_{abc} \lambda_{ab(i} \lambda_{jk)c} - 4 \lambda_{iab}\lambda_{jbc}\lambda_{kca} \,,
\ee
where the round parentheses denote a symmetrization of the enclosed indices.
This is a gradient flow in the sense that $\beta_{ijk} \delta \lambda_{ijk} = \delta A$
where the function $A$ is given by
\be 
A=-\epsilon \lambda_{ijk}\lambda_{ijk}+\frac{1}{4}\lambda_{bmn}\lambda_{imn}\lambda_{bjk}\lambda_{ijk}-\lambda_{abi}\lambda_{bcj}\lambda_{cak}\lambda_{ijk}
\ee
and at the fixed point has the value 
\be
A_*\stackrel{\mathrm{LO}}{=}-\frac{\epsilon}{8} \lambda_{*ijk}\lambda_{*ijk}\stackrel{\mathrm{LO}}{=}\, -\frac{3}{8} \epsilon \sum_i \gamma_i.
\label{Avalue}
\ee 
Another quantity of interest is the coefficient of the energy-momentum tensor two point function $C_T$, which can be related to $A$ at two-loop for any fixed point
\be
\frac{C_T}{C_{T,scalar}}=3-\frac{7}{18}  \lambda_{*ijk}\lambda_{*ijk}\stackrel{\mathrm{LO}}{=}3+\frac{28}{9\epsilon} A_*\,,
\ee
where $C_{T,scalar}$ refers to the single free scalar theory's coefficient.
These functions may be calculated perturbatively at higher orders as well \cite{Grinstein:2014xba,Gracey:2015fia,Osborn:2017ucf}, and every statement made here can be generalized beyond LO to the three-loop level.
Our goal is to find all the zeroes of the above set of beta functions. This task is significantly facilitated by choosing a more convenient basis of couplings.
%
%


\section{Choice of couplings} \label{sec:choice}

%
For general number of fields a full understanding of the theory space and fixed points is a highly nontrivial task.
Since a complete analysis of the single and double field models already exists, we concentrate here on three-flavor model, that is the lowest number of flavors which has not yet been fully explored. The kinetic term in this case is invariant under general three dimensional rotations $ U\in O(3), \; U^TU=1 $,
which are the maximal symmetry that the system can possess \cite{Osborn:2017ucf}.
The group induces a transformation on the potential $v(\phi) \to v'(\phi)=v(U\cdot\phi)$ and hence on the couplings
themselves
$ \lambda_{ijk} \rightarrow U_{ia}U_{jb}U_{kc}\lambda_{abc} $.
Notice that $O(3)$ is not necessarily a symmetry of any fixed points and therefore $v\neq v'$ in general.

For three flavors, there are ten independent couplings in terms of which the potential \eqref{v} may be expressed more explicitly as 
{\setlength\arraycolsep{2pt}
\begin{equation}
 \begin{split}
	v &= \frac{1}{6} \left(\lambda_1 \phi_1^3+3\lambda_2\phi_1^2\phi_2+3\lambda_3\phi_1^2\phi_3+3 \lambda_4 \phi_1\phi_2^2 +3 \lambda_7\phi_1\phi_3^2  \right. \nn\\
	&+ \left. 6 \lambda_5 \phi_1 \phi_2 \phi_3 +\lambda_6 \phi_2^3+\lambda_{10}\phi_3^3+3\lambda_9\phi_2 \phi_3^2+3\lambda_8 \phi_2^2 \phi_3\right)
 \end{split}
\end{equation}
Rather than analyzing the zeroes of the betas of $ \lambda_I, \; I=1,2\cdots 10$,
we find it more convenient to move to a basis where this ten-dimensional representation is manifestly split into its irreducible components $ 10 = 7 \oplus 3 $. The resulting couplings, denoted by $ g_I, \;  I=1,2\cdots 10 $, are related to the original $ \lambda_I $ couplings through the following linear combinations
\be \label{r7}
r_7 = \left(\ba{c} 5(\lambda_{10}-3\lambda_8) \\ 5(3\lambda_{9}-\lambda_6) \\ -20\lambda_5 \\ 10(\lambda_7-\lambda_4) \\ 4\lambda_3-\lambda_8-\lambda_{10} \\ 4\lambda_2-\lambda_6-\lambda_9 \\ 2\sqrt{2}(2\lambda_1-3(\lambda_4+\lambda_7))\ea\right) \equiv \left(\ba{c} g_1 \\ g_2 \\ g_3 \\ g_4 \\ g_5 \\ g_6 \\ g_7 \ea\right),
\ee
\be \label{r3}
r_3 = \left(\ba{c} \lambda_3+\lambda_8+\lambda_{10} \\ \lambda_2+\lambda_6+\lambda_9 \\ \sqrt{2}(\lambda_1+\lambda_4+\lambda_7)\ea\right) \equiv \left(\ba{c} g_8 \\ g_9 \\ g_{10} \ea\right),
\ee
where $ r_7$ and $ r_3$  carry, respectively, the seven-dimensional and the three-dimensional irreducible representations. This can also be seen through the alternative decomposition where the tensor $ \lambda_{ijk} $ is to split into a vector $ \kappa_i $ and a symmetric traceless tensor $ \sigma_{ijk} $ as follows:
\be 
\lambda_{ijk} = \kappa_{(i}\delta_{jk)}+\sigma_{ijk}, \qquad \sigma_{ill}=0\,,
\ee
where the irreps are expressed in terms of $\lambda_{ijk}$ as
\be 
\kappa_i = \frac{3}{N+2}\lambda_{ill}, \qquad
\sigma_{ijk} = \lambda_{ijk} - \frac{3}{N+2}\lambda_{ll(i}\delta_{jk)}\,,
\ee
given here for general number of flavors $N$.
For the three flavor case, one can make the identifications $ \kappa_i \leftrightarrow r_3 $ and $ \sigma_{ijk} \leftrightarrow r_7 $.

The $ g_I $ couplings in \eqref{r7} and \eqref{r3} could be further chosen such that the components of the irreps have definite values among $m=0,\pm 1, \pm 2, \pm 3$ under $J_3$, the third component of angular momentum which leaves $ \phi_3 $ untouched. However we have found more convenient to choose a slightly different basis where couplings with the same $|m|\neq 0$ value are linearly combined, in such a way that the resulting pair of couplings transform with $R(|m|\theta)$ under a rotation with angle $\theta$ around the third axis, where 
$R(\theta)$ is a simple $2\times 2$ matrix that rotates a vector clockwise by an angle $\theta$.
More explicitly, under $\exp(-i\theta J_3)$ the couplings $ g_1 $ and $ g_8 $ which have $ m=0 $ remain unaltered, while the pairs $ (g_2,g_3) $ and $ (g_9,g_{10}) $ transform with $R(\theta)$, and the pairs $ (g_4,g_5) $ and $ (g_6,g_7) $ transform, respectively, with $R(2\theta)$ and $R(3\theta)$.

The next observation that proves useful in our analysis is that there is a redundancy in the space of fixed points, that is any rotation of a fixed point is itself a fixed point. This is because the betas of the rotated couplings are the rotated betas of the original couplings, i.e.\ $ \beta(Ug) = U\beta(g) $, where $ U $ is implicitly assumed to be in the appropriate representation. This further shows that $ O(3) $ related fixed points are physically equivalent. This equivalence of fixed points under rotation gives us a freedom to constrain the space within which we are seeking for fixed points by setting some of the couplings to zero. Consider, for instance, $ r_3 $ which carries the fundamental representation of $ O(3) $. By a suitable rotation, one can always align the vector along, say, the eight direction, that is, to set $ g_9 = g_{10} =0 $. This breaks the $ O(3) $ freedom to the $ O(2) $ subgroup generated by $ J_3 $. We may then use this remaining freedom to set further constraints on the set of couplings included in $ r_7 $. For instance, we may set $ g_2=0 $ by a rotation around the $ \phi_3 $ axis. This will completely fix the redundancy and simplify the beta functions.

In order to look for fixed points, we completely remove the redundancy first, that is, we set $g_2 = g_9 = g_{10} = 0$. This will prevent finding multiple zeros of the beta functions that are equivalent. However, there will still be ten coupled betas that are functions of seven couplings and therefore difficult to solve. We have observed that omitting the three betas $\beta_2$, $\beta_9$, and $\beta_{10}$ of the redundant couplings, the zeroes of the remaining betas can always be found numerically and sometimes even analytically.
Of course, these roots are only admissible if they make the three betas $\beta_2$, $\beta_9$, and $\beta_{10}$ vanish as well, but this is a straightforward test that discards all the inconsistent roots and returns only the admissible ones.
In this way, we have been able to find all the $N=3$ fixed points of the $d=6-\epsilon$ scalar model, without imposing any extra constraint.
%


\section{Results}


%
The results of this analysis are collected in Table~\ref{fps} which includes the symmetries of the fixed points, the field anomalous dimensions, and the values of $A_*$ at LO, 
rescaled by $\epsilon$ and $\epsilon^2$, respectively.
For completeness, we have included the information about the single and two-flavor models as well (we recover the $N=2$ classification reported in \cite{Osborn:2017ucf}). Clearly, by putting fixed points together one, can construct fixed points with higher number of flavors. To avoid such trivial cases, here we have restricted ourselves only to fully interacting (irreducible) fixed points, for which there is no basis where there are two decoupled sectors. For three scalars, there are six fully interacting fixed points with real anomalous dimensions altogether. 
The fixed point with $O(2)$ symmetry belongs to a known family pointed out in \cite{Fei:2014yja}.
Two fixed points have permutation and ${\mathrm{PT}}$ symmetry: $ S_4\times \mathbb{Z}_2^{\mathrm{PT}} $ and  $ S_3\times \mathbb{Z}_2^{\mathrm{PT}} $
(we refer to appendix~\ref{sect:appendix-further} for more information on ${\mathrm{PT}}$ symmetry.)
The last three fixed points are such that the three anomalous dimensions are all different.
One has the symmetry of the Klein four-group
$\mathbb{K}_4=\mathbb{Z}_2 \times \mathbb{Z}_2$.
This is particularly interesting because the field anomalous dimensions are also all positive, which suggests that the theory is perturbatively unitary, such as the $O(2)$ symmetric fixed point.
\renewcommand{\arraystretch}{1.55}
\begin{center}
	\begin{table} 
		\begin{tabular}{|c|c|l|r|} 
			\hline
			$N$ & Anomalous dimensions$/\epsilon$ & Symmetry & A$ _*/\epsilon^2$   \\ \hline \hline
			1 & $-\frac{1}{18}$ & $\mathbb{Z}_2^{\mathrm{PT}} $ & 0.02083  \\ \hline \hline
			\multirow{2}{*}{2} & $(\frac{1}{6},\frac{1}{6})$  & $S_3$ & -0.125  \\ \cline{2-4}
			& $(-\frac{61}{998},-\frac{25}{499})$  & $\mathbb{Z}_2\times \mathbb{Z}_2^{\mathrm{PT}}$ & 0.04171   \\ \hline \hline
			\multirow{6}{*}{3} & $(-\frac{1}{6},-\frac{1}{6},-\frac{1}{6})$  & $S_4\times \mathbb{Z}_2^{\mathrm{PT}}$ & 0.1875  \\ \cline{2-4}
			& $(0.093267,0.093267,0.167418 )$  & $O(2)$ & -0.1327 \\ \cline{2-4}
			& $(-\frac{401}{7994},-\frac{401}{7994},-\frac{533}{7994})$  & $S_3\times \mathbb{Z}_2^{\mathrm{PT}}$ & 0.06263   \\ \cline{2-4}
			& $(\frac{157+3 \sqrt{561}}{1698},\frac{157-3 \sqrt{561}}{1698},\frac{289}{1698})$  & $\mathbb{K}_4$ & -0.1332   \\ \cline{2-4}
			& $(0.168983,0.1653200,-0.059256)$  & $\mathbb{Z}_2\times \mathbb{Z}_2^{\mathrm{PT}}$ & -0.1031 \\  \cline{2-4}
			& $(-0.063434,-0.055612,-0.047844)$  & $\mathbb{Z}_2 \times \mathbb{Z}_2^{\mathrm{PT}}$ & 0.06258 \\ \hline
		\end{tabular}
		\caption{Fixed points with $N=1,2,3$ with their symmetries, anomalous dimensions and the values of $A$ at LO.} \label{fps}
	\end{table}
\end{center}
\noindent

For three fields, the fully interacting fixed point potentials (rescaled by $\sqrt{\epsilon}$) at LO in the particular basis described in the previous section are the following
(in the same order as in Table~\ref{fps} from top to bottom):
{\small 
\begin{eqnarray*}
V_1 &=& \frac{i}{2} \phi_1 \phi_2 \phi_3  \\
V_2 &=& a \, \phi_3^3 +b\, \phi_3( \phi_1^2+ \phi_2^2)  \\
V_3 &=& i \frac{ \left(13 \sqrt{26}\, \phi _3^3+30 \sqrt{26} \left(\phi _1^2\!+\!\phi _2^2\right) \phi _3\!-\!\sqrt{1009}
\! \left(\phi _1^2\!-\!3 \phi _2^2\right)\phi _1\right)}{12 \sqrt{11991}} \\
V_4 &=& \frac{\phi _3 \left(4 \sqrt{33} \, \phi _3^2\!+\!3 \left(\sqrt{17}\!-\!3 \sqrt{33}\right) \phi _2^2\!-\!3 \left(\sqrt{17}\!+\!3
\sqrt{33}\right) \phi _1^2\right)}{12 \sqrt{566}}  \\
V_5 &=& \phi_2 \left( a_5 \phi _1^2+b_5 \phi_2^2+c_5\phi_3^2\right) + i\,\phi_3 \left( d_5 \phi _1^2+e_5 \phi_2^2+f_5\phi_3^2\right) \\ 
V_6 &=& i \left[ \phi_2 \left( a_6 \phi _1^2+b_6 \phi_2^2+c_6\phi_3^2\right) + \phi_3 \left( d_6 \phi _1^2+e_6 \phi_2^2+f_6 \phi_3^2\right)\right] \,
\end{eqnarray*}
 }%
\vskip 0.1cm
\noindent where the coefficients $ a,b $ which can also be calculated analytically are 
{\small
\be 
a = -0.0786083,\qquad \qquad b=0.187016 \nn\,,
\ee
}%
while the remaining twelve coefficients which have been calculated only numerically are as follows:
{\small
{\setlength\arraycolsep{7pt}
\be  
\ba{lll}
a_5=0.258788   & b_5= -0.0909904 & c_5= 0.0141828 \nn\\
d_5= -0.0705451 & e_5=0.0615644 & f_5= -0.0694790 \nn\\
a_6= 0.078108 & b_6= -0.0228088 & c_6= -0.0096818 \nn\\
d_6= 0.121464 & e_6=0.112409 & f_6= 0.046825 \,.\nn
\ea
\ee}%
}%
From Table~\ref{fps}, one can make the observation that (at fixed number of flavors),
among the real fixed points and, separately, among the purely imaginary fixed points the larger is the symmetry the bigger is the value of the $ A $ function at the fixed point.
(i.e.\ leaving aside the truly complex one).
We note that these statements include decomposable fixed points, i.e.\ those that are not fully interacting such as three copies of the Lee-Yang model, but excludes fixed points with Gaussian factors.

For each critical theory, we have computed with next-to-leading order accuracy the critical exponents $\theta_{2,j}$
of the six relevant and $\theta_{3,k}$ of the ten marginal couplings as the negative of the eigenvalues of the RG stability matrix at the fixed point.
The scaling dimensions of the corresponding operators can be obtained using the relation $\Delta_i=d-\theta_i$ except for the three quadratic operators corresponding to the equations of motion ${\cal O}\propto \partial_{\phi_i}V$, because they satisfy the scaling relations $d-2+2\gamma_i=2\theta_{2,i}$ and are descendants in the sense of CFT
\cite{Codello:2017qek,Codello:2018nbe,Codello:2017hhh}.
Concentrating our attention on the two real fixed points $O(2)$ and $\mathbb{K}_4$,
we notice that for small $\epsilon$ the first has one more relevant direction than the second in the cubic sector, making $\mathbb{K}_4$ \emph{more infrared stable}.
Neglecting the equations of motion, in the quadratic sector both points have a singlet operator which is traditionally
associated with the scaling of the correlation length and hence the exponent $\nu$.
The $O(2)$ point has also two operators which raise/lower the $U(1)\simeq O(2)$ charge,
while the $\mathbb{K}_4$ point has an operator which respects the symmetry but gives contributions with different signs to either mass of two components, 
and another ($\propto \phi_1 \phi_2$ at LO) which is responsible for the breaking pattern $\mathbb{K}_4\to \mathbb{Z}_2$.
We summarize all information in Tables~\ref{summary}
~\ref{nlo-crits} and~\ref{nlo-crits-supp}.
\renewcommand{\arraystretch}{1.55}
\begin{center}
	\begin{table} 
		\begin{tabular}{|c|c|c|} 
			\hline
            exp.\ & $O(2)$ & $\mathbb{K}_4$ \\
			\hline\hline
			${\rm dim}(\theta_3>0)$ & 9 & 8 \\
			\hline\hline
			\multirow{3}{*}{$\gamma$} & ${\small 0.09327\epsilon+0.17241 \epsilon^2}$  & ${\small 0.1343\epsilon + 1.4033 \epsilon^2}$  \\
            & ${\small 0.16742\epsilon+0.30275 \epsilon^2}$ & ${\small 0.0506\epsilon - 1.0727 \epsilon^2}$  \\
            &   & ${\small 0.1702\epsilon + 0.5280 \epsilon^2}$ \\
			\hline \hline
			\multirow{1}{*}{$\theta_2=\nu^{-1}$} & ${\small 2+1.2606\epsilon-0.0833\epsilon^2}$  & ${\small 2+1.4115 \epsilon - 0.0419 \epsilon^2}$  \\
			\hline \hline
			\multirow{4}{*}{$\theta_2$ break } & ${\small 2+0.3731 \epsilon+0.2268\epsilon^2}$  & ${\small 2+0.3098 \epsilon + 0.1738\epsilon^2}$  \\
			& (charge operator)  & ($\mathbb{K}_4\to \mathbb{Z}_2$) \\
			& & ${\small 2+0.2504 \epsilon + 0.1764 \epsilon^2}$\\
			& & ($m_1^2/m_2^2<0$) \\
			\hline 
		\end{tabular}
		\caption{Summary of the most important properties of the real fixed points to NLO.} \label{summary}
	\end{table}
\end{center}
\noindent

\renewcommand{\arraystretch}{1.55}
\begin{table} 
\begin{equation} 
\begin{array}{|c|c|c|}
\hline
\mathrm{Symm.} & O(2) & \mathbb{K}_4 
\\ \hline
& {\small 0.09327\epsilon+0.17241 \epsilon^2}  & {\small 0.1343\epsilon + 1.4033 \epsilon^2}  \\
\cline{2-3}
\gamma & {\small 0.09327\epsilon+0.17241 \epsilon^2}   & {\small 0.0506\epsilon - 1.0727 \epsilon^2}   \\ \cline{2-3}
& {\small 0.16742\epsilon+0.30275 \epsilon^2} & {\small 0.1702\epsilon + 0.5280 \epsilon^2}  \\ \hline
& {\small 3.0762\epsilon+1.9630 \epsilon^2}  & {\small 3.2831\epsilon + 2.3102\epsilon^2}  \\ \cline{2-3}
& {\small 2.3566\epsilon+0.5370 \epsilon^2} & {\small 3.0485\epsilon + 1.6481 \epsilon^2}   \\ \cline{2-3}
& {\small 2.3566\epsilon+0.5370 \epsilon^2} & {\small 2.3052\epsilon + 0.5648 \epsilon^2}   \\ \cline{2-3}
& {\small 1.8990\epsilon+0.5943 \epsilon^2} & {\small 1.5753\epsilon + 0.2835\epsilon^2}   \\ \cline{2-3}
\theta_3 & {\small 1.8990\epsilon+0.5943 \epsilon^2} & {\small 1.3701\epsilon + 0.1004\epsilon^2}   \\ \cline{2-3}
& {\small -\epsilon+5.9419 \epsilon^2}  & {\small -\epsilon + 7.4594 \epsilon^2}  \\ \cline{2-3}
& {\small 0.0405\epsilon+3.3479 \epsilon^2} & {\small -0.0664\epsilon + 2.7930 \epsilon^2}  \\ \cline{2-3}
& {\small 0.0405\epsilon+3.3479 \epsilon^2} & {\small 3.3485 \epsilon^2}  \\ \cline{2-3}
& {\small 1.2131\epsilon^2} & {\small 1.6826 \epsilon^2}   \\ \cline{2-3}
& {\small 0.3768\epsilon^2}  & {\small 1.1651 \epsilon^2}   \\ \hline
& {\small 2+1.2606\epsilon-0.0833\epsilon^2} & {\small 2+1.4115 \epsilon - 0.0419 \epsilon^2}  \\ \cline{2-3}
& {\small 2-0.4067\epsilon+1.0623\epsilon^2}  & {\small 2+0.2504 \epsilon + 0.1764 \epsilon^2}  \\ \cline{2-3}
\theta_2 & {\small 2-0.4067\epsilon+1.0623\epsilon^2}  & {\small 2-0.4494 \epsilon + 0.6501\epsilon^2}   \\ \cline{2-3}
& {\small 2+0.3731 \epsilon+0.2268\epsilon^2} & {\small 2-0.3657 \epsilon + 1.6288\epsilon^2}   \\ \cline{2-3}
& {\small 2+0.3731 \epsilon+0.2268\epsilon^2} & {\small 2+0.3098 \epsilon + 0.1738\epsilon^2}    \\ \cline{2-3}
& {\small 2-0.3326\epsilon + 1.1827\epsilon^2} & {\small 2-0.3298 \epsilon + 1.4700\epsilon^2}  \\ \hline
\end{array} \nn
\end{equation}
\caption{NLO results for the coupling scaling dimensions of the two perturbatively unitary critical theories.} \label{nlo-crits}
\end{table} 
\noindent

Moreover, it may be useful to report some structure constants
for these critical models, considered as CFTs, as well. 
They can be calculated at LO in several ways. Here, we follow 
\cite{Codello:2018nbe}
and give as an example two structure constants for the fixed point $\mathbb{K}_4$, %
\be \label{ope-o2}
C_{\phi_3\phi_3\mathcal{S}_3} = \frac{289}{3 \sqrt{56883}} \, \epsilon
\, , \quad
C_{\phi_3\mathcal{S}_2\mathcal{S}_2} = 3 \sqrt{\frac{66}{283}} \,\epsilon ,
\ee
where $ \mathcal{S}_3 $ is the cubic scaling operator corresponding
to the LO scaling dimension $ -\epsilon $,
and $ \mathcal{S}_2=\phi_1\phi_2$ is the quadratic scaling operator
responsible of $\mathbb{K}_4 \to \mathbb{Z}_2$ breaking.
The same structure constants can be calculated for the $ O(2) $ fixed point which turn out to be
\be \label{ope-z22}
C_{\phi_3\phi_3\mathcal{S}_3} = -0.397965 \, \epsilon 
\, , \quad
C_{\phi_3\mathcal{S}_2\mathcal{S}_2} =  -1.49613 \,  \epsilon \,.
\ee
The explicit form of these operators can be found in appendix~\ref{sect:appendix-betas}.
Another simple example of a structure constant is $ C_{\phi_i\phi_j\phi_j} = -2\, v_{ijk} $ which is proportional to $\sqrt{\epsilon}$. 
This means, for instance, that $ C_{\phi_1\phi_2\phi_3} $ is nonzero only for the $S_4\times \mathbb{Z}_2^{\mathrm{PT}}$  invariant fixed point. 
Finally we also give the two-loop information for the $C_T$ of these two fixed points,
\be
\frac{C_T^{O(2)}}{C_{T,scalar}}=3-0.4128\, \epsilon \,, \quad  \quad  \frac{C_T^{\mathbb{K}_4}}{C_{T,scalar}}=3-\frac{469}{1132} \epsilon \,. \nn
\ee
%


\section{Outlook}


The combination of perturbative RG and $\epsilon$-expansion is unequivocally
a fundamental tool to investigate interacting scale invariant QFTs,
which we used to discuss the general quest of finding, without any prior assumption on symmetry,
all possible inequivalent fixed points of the RG, here interpreted as universality classes of critical phenomena.
We argued that, in the absence of any symmetry, the task of finding all solutions of RG equations
becomes a complicate algebraic problem which grows very rapidly with the number $N$ of order parameters.
Nevertheless, it is a fundamental step to undertake if one desires to
elucidate the global structure of the theory space of $N$-components fields near the critical dimension
or even for arbitrary $d>2$.

It can be difficult to fully realize the central role that symmetry has in simplifying almost all field theory's results,
until one tries without that.
By all means, even if we partly circumvented the absence of symmetry with clever application of the irreducible representations
of a maximal group, 
our computations have been rather difficult precisely because of the lack of a symmetry structure to begin with.
The payoff is very big: we could see very generally how symmetries emerge as a property of the universality classes,
here understood as a fixed point of the renormalization group, rather than as ingredients.

To elucidate our point of view and the method, we have focused on systems with three scalar order parameters in $d=6-\epsilon$
and adopted a convenient group theoretical basis for the coupling's space which allowed us,
by removing the parametrization redundancy, to deal with a tractable algebraic problem.
The main outcome of our investigation is the discovery of six fully interacting fixed points with real critical exponents.
With the exception of a fixed point with $O(2)$ symmetry which was already known in the literature,
all other five fixed points are completely new.
The most interesting fixed point has $\mathbb{K}_4=\mathbb{Z}_2\times\mathbb{Z}_2$ symmetry,
and is perturbatively unitary as displayed by three different field's anomalous dimensions.
All the properties of this latter fixed point are rather unique; since it is reasonable to assume that its existence
might continue down to lower dimensions, we hope that our findings might stimulate an independent search using
numerical conformal bootstrap.

All other new critical theories exhibit a combination of discrete and PT symmetry.
The main properties of all FPs are summarized in Tables~\ref{fps},~\ref{summary},~\ref{nlo-crits} and~\ref{nlo-crits-supp}.
They include the analysis of all the quadratic and cubic scaling operators, the latter related to the stability under the RG flow of these fixed points.

It is worth emphasizing that these solutions are the only ones we could find, using a combination of
analytical and numerical methods, that solve the fixed point equations.
We are rather confident that no further solution will ever emerge; therefore, our work has heuristically completed
the classification of $N=3$ critical field theories in $d=6-\epsilon$.
We could expect them to be the only ones that could survive the continuation to finite values of $\epsilon$
and thus exist in dimension five, four, or three (if higher derivative interactions are excluded~\cite{SafariVacca}),
but the statement might change because of nonperturbative effects.
The knowledge of the fixed points in the $\epsilon$-expansion and their symmetries provides a starting point for further theoretical investigations in lower dimensions with alternative approaches such as conformal bootstrap~\cite{ElShowk:2012ht}, nonperturbative RG ~\cite{Polchinski:1983gv,Wetterich:1992yh,Morris:1994ie}
or lattice Monte Carlo methods \cite{Pelissetto:2000ek}, 
but also provides suggestions for possible experimental realizations.
Not all critical theories present at small $\epsilon$ necessarily extend to integer dimensions less than the critical one, 
but if some exist they will maintain the symmetries found by our general analysis.
Furthermore, a future complete nonperturbative analysis is necessary to determine how the found fixed points are connected by RG flow trajectories and therefore understand which ones are ultraviolet or infrared with respect to each other.
Finally, we have found several perturbatively nonunitary critical theories with complex scaling dimensions (which we do not report) that can in principle constitute realizations of complex CFTs \cite{Gorbenko:2018ncu}.

%

\appendix

\begin{widetext}

\section{Beta functions and additional results}\label{sect:appendix-betas}

The explicit form of the beta functions of the couplings in the irreducible basis which we introduced in the main text is at the leading one-loop order, after imposing $g_2 = g_9 = g_{10} = 0$, as described in the main text,
\begin{eqnarray*}
 \beta_1 & = & -\frac{\epsilon}{2}g_1 +\frac{1}{400} \left(g_1^3+2 g_1 \big(6 g_3^2+3g_4^2+3g_5^2-2g_6^2-2g_7^2\big)+12 \sqrt{2} g_3\left( g_4 g_7-g_5 g_6-g_3 g_4\right)\right)\nn\\
 &&+\frac{1}{25} \left(3 g_1^2+17 g_3^2-2 g_4^2-2 g_5^2-g_6^2-g_7^2\right) g_8 -\frac{43}{50} g_1 g_8^2 + \frac{4}{5}g_8^3
\\
 \beta_2 &=& -\frac{1}{400} \left((12 g_3^2+g_4^2-g_5^2) g_6+2 g_4 g_5 g_7-\sqrt{2} g_1 \left(g_4 g_6+g_5 \left(2 g_3+g_7\right)\right)\right)
 +\frac{1}{75 \sqrt{2}} g_8\left(g_4 g_6+g_5 \left(g_7-18 g_3\right)\right)
\\
 \beta_3 &=& -\frac{\epsilon}{2}g_3 +\frac{1}{400} \left(22 g_3^3-12 g_7 g_3^2+g_1^2 g_3+(g_4^2+g_5^2-2 (g_6^2+g_7^2)) g_3 -2 g_4 g_5 g_6+(g_4^2-g_5^2) g_7 \right.
 \nn\\&&
 \left.+\sqrt{2} g_1 (g_4 (g_7-2 g_3)-g_5 g_6)\right)  
 +\frac{1}{150}  g_8 \big(17 g_1 g_3+\sqrt{2} (18g_4 g_3+g_4g_7-g_5 g_6)\big) -\frac{59}{75} g_3 g_8^2
\\
 \beta_4 &=& -\frac{\epsilon}{2}g_4 +\frac{1}{400} \left(-2 g_4^3+\big(5 g_1^2+2 (5 g_3^2+10 g_7 g_3-g_5^2+g_6^2+g_7^2)\big) g_4
 -10 g_3 \big(2 g_5 g_6+\sqrt{2} g_1 (g_3-g_7)\big)\right) \nn\\
 && +\frac{1}{15} g_8\left(\sqrt{2} g_3 \left(9 g_3+g_7\right)-2 g_1 g_4\right)  -\frac{17}{30} g_4 g_8^2
\\
 \beta_5 &=& -\frac{\epsilon}{2}g_5 +\frac{1}{400} \left(-2 g_5^3+g_5\big(5 g_1^2+2 (5 g_3^2-10 g_7 g_3-g_4^2+g_6^2+g_7^2)\big)-10 g_3 (\sqrt{2} g_1+2 g_4) g_6\right) \nn\\
 && -\frac{1}{15} \left(2 g_1 g_5+\sqrt{2} g_3 g_6\right) g_8 -\frac{17}{30} g_5 g_8^2
\\
 \beta_6 &=& -\frac{\epsilon}{2}g_6 +\frac{1}{400} \left(2 g_6^3+\big(-5 g_1^2-30 g_3^2+2 g_7^2+3(g_4^2+g_5^2)\big) g_6-15 g_3 (\sqrt{2} g_1+2 g_4) g_5\right) \nn\\
 && -\frac{1}{10} \left(\sqrt{2} g_3 g_5+g_1 g_6\right) g_8 -\frac{1}{5} g_6 g_8^2
 \\
 \beta_7 &=& -\frac{\epsilon}{2}g_7 +\frac{1}{400} \left(2 g_7^3+\big(-5 g_1^2-30 g_3^2+2 g_6^2+3 (g_4^2+g_5^2)\big) g_7  -15 g_3 \big(4 g_3^2+g_5^2-g_4 (\sqrt{2} g_1+g_4)\big)\right) \nn\\
 &&  + \frac{1}{10} \left(\sqrt{2} g_3 g_4-g_1 g_7\right) g_8 -\frac{1}{5} g_7 g_8^2
\\
 \beta_8 &=& -\frac{\epsilon}{2}g_8 +\frac{1}{600} \left(g_1^3+(17 g_3^2-2 g_4^2-2 g_5^2-g_6^2-g_7^2) g_1+2 \sqrt{2} g_3 \left(g_4 \left(9 g_3+g_7\right)-g_5 g_6\right)\right) \nn\\
 && -\frac{1}{1200} \, g_8 \left(43 g_1^2+472 g_3^2+34 g_4^2+34 g_5^2+8 (g_6^2+g_7^2)\right) + \frac{1}{10} g_1 g_8^2 -\frac{21 g_8^3}{50}
\\
 \beta_9 &=& \frac{1}{1200}\left(2(8 g_3^2-g_4^2+g_5^2) g_6-4 g_4 g_5 g_7-\sqrt{2} g_1 \left(3 g_4 g_6+g_5 \left(g_3+3 g_7\right)\right)\right)
 -\frac{11}{600 \sqrt{2}} g_8\left(g_4 g_6+g_5 \left(g_7-3 g_3\right)\right)
\\
 \beta_{10} &=& \frac{1}{1200}\left(24 g_3^3+16 g_7 g_3^2+2 g_1^2 g_3+\big(6 g_4^2+6 g_5^2-8 (g_6^2+g_7^2)\big) g_3  -4 g_4 g_5 g_6+\sqrt{2} g_1 \left(3 g_5 g_6+g_4 \left(g_3-3 g_7\right)\right)
 \right.
 \nn\\&&
 \left.
 +2 (g_4^2-g_5^2) g_7\right)
 -\frac{11}{1200}\, g_8 \left(2 g_1 g_3+\sqrt{2} \left(g_4 \left(3 g_3+g_7\right)-g_5 g_6\right)\right) + \frac{1}{5} g_3 g_8^2\,.
\end{eqnarray*}

\end{widetext}

The above beta functions can be derived combining the beta functional of the potential and the anomalous dimension matrix.
For the derivation of the results on the various critical exponents reported in Tables~\ref{nlo-crits} and Table~\ref{nlo-crits-supp} we have actually used
the NLO expansion to investigate possible degeneracies of fixed points.
The NLO beta functional, anomalous dimension matrix, and flow of the general $\lambda_{ijk}$ couplings
are

\begin{widetext}
\begin{eqnarray*}
 \beta_v &=& -d v +\frac{d-2}{2}\phi_i v_i + \phi_i \gamma_{ij} v_j - \frac{2}{3} v_{ij}v _{jk} v_{ki}
 -2\, v_{im}v_{mj}v_{kl}v_{ikn}v_{jln} 
 +\frac{7}{9} v_{ij}v_{jk}v_{kl}v_{abi}v_{abl}
 - \frac{4}{3} v_{il}v_{jm}v_{kn}v_{ijk}v_{lmn} 
 \\
 \gamma_{ij}
 &=& \frac{1}{3} \lambda_{iab} \lambda_{jab}\;  +\,\frac{8}{9} \lambda_{ikl} \lambda_{jpq}\lambda_{kpm} \lambda_{lqm}-\frac{11}{27}\lambda_{imk} \lambda_{jml}\lambda_{pqk} \lambda_{pql}
 \\
 \beta_{ijk}
 &=& -\frac{1}{2}\epsilon\, \lambda_{ijk} + \lambda_{abc} \lambda_{ab(i} \lambda_{jk)c} - 4 \lambda_{iab}\lambda_{jbc}\lambda_{kca}
 +\,\frac{8}{3} \lambda_{rpm} \lambda_{sqm}\lambda_{rsa} \lambda_{pq(i}\lambda_{jk)a}-\frac{11}{9}\lambda_{pqr} \lambda_{pqs}\lambda_{mra} \lambda_{ms(i}\lambda_{jk)a}
 \nn\\ &&
 -12\, \lambda_{ma(i}\lambda_{j|an|}\lambda_{k)pq}\lambda_{mpl}\lambda_{nql} +\frac{14}{3} \lambda_{mp(i}\lambda_{|pq|j}\lambda_{k)qn}\lambda_{abm}\lambda_{abn}
 -8 \lambda_{ial}\lambda_{jbm}\lambda_{kcn}\lambda_{abc}\lambda_{lmn}
\end{eqnarray*}
\begin{table} \label{table3}
	\begin{tabular}{|c|c|c|c|c|}
		\hline
		Symm. & $S_4\times \mathbb{Z}_2^{\mathrm{PT}}$  & $S_3\times \mathbb{Z}_2^{\mathrm{PT}}$ & $\mathbb{Z}_2\times \mathbb{Z}_2^{\mathrm{PT}}$ & $\mathbb{Z}_2\times \mathbb{Z}_2^{\mathrm{PT}}$ \\ \hline
		\multirow{3}{*}{$\gamma$} & $-\frac{\epsilon}{6}-\frac{10\epsilon^2}{27}$  &  $-\frac{401\epsilon}{7994}-\frac{1511230848\epsilon^2}{63856107973}$ & {\small $0.1690\epsilon + 0.6721 \epsilon^2$} & {\small $-0.0634\epsilon - 0.0386 \epsilon^2$}   \\ \cline{2-5} 
		& $-\frac{\epsilon}{6}-\frac{10\epsilon^2}{27}$  & $-\frac{401\epsilon}{7994}-\frac{1511230848\epsilon^2}{63856107973}$ &  {\small $0.1653\epsilon + 0.6480 \epsilon^2$} & {\small $-0.0556\epsilon - 0.0297 \epsilon^2$}   \\ \cline{2-5}
		& $-\frac{\epsilon}{6}-\frac{10\epsilon^2}{27}$  & $-\frac{533\epsilon}{7994}-\frac{5387208471\epsilon^2}{127712215946}$  &  {\small $-0.0593\epsilon - 0.0536 \epsilon^2$} & {\small $-0.0478\epsilon - 0.0209 \epsilon^2$} \\ \hline
		\multirow{10}{*}{$\theta_3$} & $-\frac{(1+\sqrt{265})\epsilon}{6}+\frac{839 \epsilon ^2}{36 \sqrt{265}}+\frac{107 \epsilon ^2}{18}$  &  {\small $-1.0045\epsilon + 1.9395\epsilon^2$} & {\small $3.6395\epsilon + 2.8931\epsilon^2$} & {\small $-1.0015\epsilon + 1.9338 \epsilon^2$}  \\ \cline{2-5}
		& $-\frac{(1+\sqrt{265})\epsilon}{6}+\frac{839 \epsilon ^2}{36 \sqrt{265}}+\frac{107 \epsilon ^2}{18}$  & {\small $-1.0045\epsilon + 1.9395\epsilon^2$} & {\small $3.6225\epsilon + 2.8686 \epsilon^2$} & {\small $-1.0045\epsilon + 1.9382 \epsilon^2$}  \\ \cline{2-5}
		& $-\frac{(1+\sqrt{265})\epsilon}{6}+\frac{839 \epsilon ^2}{36 \sqrt{265}}+\frac{107 \epsilon ^2}{18}$  & {\small $0.4918\epsilon + 0.0054\epsilon^2$} & {\small $2.3541\epsilon + 0.3075 \epsilon^2$} & {\small $-\epsilon + 1.9343 \epsilon^2$} \\ \cline{2-5}
		& $-\frac{(1-\sqrt{265})\epsilon}{6}-\frac{839 \epsilon ^2}{36 \sqrt{265}}+\frac{107 \epsilon ^2}{18}$  & {\small $-0.1224\epsilon+0.5805\epsilon^2$} & {\small $-1.0244\epsilon + 0.1093 \epsilon^2$} & {\small $0.5473\epsilon - 0.0138 \epsilon^2$}   \\ \cline{2-5}
		& $-\frac{(1-\sqrt{265})\epsilon}{6}-\frac{839 \epsilon ^2}{36 \sqrt{265}}+\frac{107 \epsilon ^2}{18}$  &  {\small $0.1040\epsilon + 0.5463 \epsilon^2$}  & {\small $-\epsilon + 13.2441 \epsilon^2$} & {\small $-0.1108\epsilon + 0.5584 \epsilon^2$} \\ \cline{2-5}
		& $-\frac{(1-\sqrt{265})\epsilon}{6}-\frac{839 \epsilon ^2}{36 \sqrt{265}}+\frac{107 \epsilon ^2}{18}$  &  {\small $0.1040\epsilon + 0.5463 \epsilon^2$} & {\small $-0.5540\epsilon + 0.4550 \epsilon^2$} & {\small $0.1107\epsilon + 0.3595 \epsilon^2$}  \\ \cline{2-5}
		& $-\epsilon+\frac{205\epsilon^2}{36}$  &  {\small $-\epsilon + 1.9369 \epsilon^2$}  & {\small $0.0482\epsilon - 2.6085 \epsilon^2$} & {\small $-0.1094\epsilon + 0.5563 \epsilon^2$}  \\ \cline{2-5}
		& $\frac{41\epsilon^2}{36}$  &  {\small $0.3815\epsilon^2$} & {\small $5.7601 \epsilon^2$} & {\small $0.3805 \epsilon^2$} \\ \cline{2-5}
		& $\frac{41\epsilon^2}{36}$  &  {\small $0.3815\epsilon^2$} & {\small $2.4290 \epsilon^2$} & {\small $0.3865 \epsilon^2$} \\ \cline{2-5}
		& $\frac{41\epsilon^2}{36}$  &  {\small $1.2131\epsilon^2$} & {\small $2.3103 \epsilon^2$} & {\small $0.3793 \epsilon^2$} \\ \hline
		\multirow{6}{*}{$\theta_2$} & $2-\frac {5\epsilon}{3} + \frac{49\epsilon^2}{27}$ & {\small $2-0.5667\epsilon + 0.4098 \epsilon^2$} & {\small $2+1.6474 \epsilon + 0.0251 \epsilon^2$} & {\small $2-0.5634 \epsilon + 0.4067 \epsilon^2$} \\ \cline{2-5}
		& $2+\frac{4\epsilon}{3}+\frac{71\epsilon^2}{54}$  & {\small $2-0.5502 \epsilon + 0.3951 \epsilon^2$} & {\small $2-0.5593 \epsilon + 0.4383 \epsilon^2$} & {\small $2-0.5556 \epsilon + 0.3996 \epsilon^2$}  \\ \cline{2-5}
		& $2+\frac{4\epsilon}{3}+\frac{71\epsilon^2}{54}$  &  {\small $2-0.5502 \epsilon + 0.3951 \epsilon^2$} & {\small $2-0.3511 \epsilon + 0.2610 \epsilon^2$} & {\small $2-0.5478\epsilon +0.3928 \epsilon^2$}    \\ \cline{2-5}
		& $2-\frac {2\epsilon} {3} + \frac {31\epsilon^2} {27}$  &  {\small $2+0.0518 \epsilon - 0.0161\epsilon^2$}  & {\small $2-0.3347 \epsilon + 2.1721 \epsilon^2$} & {\small $2+0.1117\epsilon -0.0271 \epsilon^2$}   \\ \cline{2-5}
		& $2-\frac {2\epsilon} {3} + \frac {31\epsilon^2} {27}$  & {\small $2+0.0518 \epsilon - 0.0161\epsilon^2$}  & {\small $2-0.3310 \epsilon + 2.1341 \epsilon^2$} & {\small $2+0.0506 \epsilon - 0.0159 \epsilon^2$} \\ \cline{2-5}
		& $2-\frac {2\epsilon} {3} + \frac {31\epsilon^2} {27}$  & {\small $2+0.0488\epsilon - 0.0160\epsilon^2$} & {\small $2+0.1006 \epsilon + 0.0618 \epsilon^2$} & {\small $2+0.0497 \epsilon - 0.0155 \epsilon^2$}  \\ \hline
	\end{tabular}
	\caption{Anomalous dimensions $\gamma$ and coupling dimensions of cubic $\theta_3$ and quadratic $\theta_2$ operators for the four fully interacting fixed points with $N=3$ which are not reported in Table~\ref{nlo-crits}.} \label{nlo-crits-supp}
\end{table}
\end{widetext}

Finally, the explicit form of the two scaling operators $ \mathcal{S}_3 $ and $ \mathcal{S}_2 $ used in the structure constants (OPE coefficients) \eqref{ope-o2} for the $ \mathbb{K}_4 $ fixed point and \eqref{ope-z22} for the $ O(2) $ fixed point is given, respectively, by 

\begin{widetext}

\be 
\ba{lll}
  \mathbb{K}_4 :\qquad & \displaystyle\qquad\mathcal{S}_3 = \frac{\phi _3}{6 \sqrt{1474}}
 \left(-\left(\sqrt{561}+99\right) \phi _1^2+\left(\sqrt{561}-99\right) \phi _2^2+44 \phi _3^2\right), \qquad & \qquad\mathcal{S}_2 = \phi_1\phi_2 \\[4mm]
 O(2): \qquad & \qquad\mathcal{S}_3 = 0.186858 \, \phi _3\left(\phi _3^2
 -0.444552 \left(\phi _1^2 +\phi _2^2\right) \right), \qquad & \qquad\mathcal{S}_2 = \phi_1\phi_2 \,. \nn
\ea
\ee

\end{widetext}

\section{$O(N)$ transformations, fixed point moduli, and PT symmetry}\label{sect:appendix-further}

Let us generalize this discussion to the case of arbitrary number of scalar flavors $N$ and temporarily neglect PT transformations.
Given a fixed point solution $v(\phi)$ of $\beta_v=0$, its symmetry content is defined as the subgroup $G$ of $O(N)$
which leaves it invariant. In other words, for $U\in G$, the action
$$
 U: v(\phi) \to v'(\phi)= v(U\cdot\phi)
$$
is such that $v(\phi)=v'(\phi)$. The symmetry group $G$ is a subgroup of $O(N)$ because $O(N)$ is the maximal symmetry that the model
can have. Using $G$ we can evince the structure of the manifold of equivalent fixed points for each symmetry content:
if $v(\phi)$ is a solution, then also a rotation of $v(\phi)$ is, but the two are physically distinct only if we are acting with a rotation which is not already in the symmetry subgroup $G$.
It is easy to see that the \emph{moduli} of equivalent fixed points for each universality class is isomorphic to
$$
 {\rm FP \,\, moduli}   ~ \simeq  ~  O(N)/\mathord G  ~ .
$$
For example, in the case $N=1$ we have only one solution (the Lee-Yang model) which is left invariant only by the identity; therefore, its
moduli of fixed points is $O(1)/\mathord \{1\}\simeq \mathbb{Z}_2$, implying that there are always two distinct, but physically equivalent, solutions
$i g \phi^3$ and $-i g \phi^3$.

Notice that complex solutions are still protected by symmetry: in the above example of the Lee-Yang model we could move from one solution to the other
by either parity or complex conjugation so the two combined leave the solution invariant. For this reason we define the PT transformation
which goes \emph{beyond} $O(N)$ by acting
$${\rm PT}:v(\phi)\to v^*({\rm P}\cdot \phi)$$
in which $v^*$ is the complex conjugate of $v$ and ${\rm P}\in O(N)$ acts on $\phi^i$ by flipping the sign of one specific field component.
This new symmetry protects potentials with complex factors and ensures that their spectra are bounded from below inasmuch those of the purely real models.
We indicated the presence of PT symmetry in the solutions by including $\mathbb{Z}^{\rm PT}_2$ in their symmetry factors;
however, it is important to drop these factors when applying the above coset formula for the moduli.
Going back once more to the $N=1$ example of the Lee-Yang model, it is trivial to see that there can be only one parity and PT maps $v(\phi) \to v^*(-\phi)=v(\phi)$
and therefore is an extension of the original symmetry.


\end{document}